# Ultrathin Oxide Films by Atomic Layer Deposition on Graphene

Luda Wang[1], Jonathan J. Travis[2], Andrew S. Cavanagh[2], Xinghui Liu[1], Steven P. Koenig[1], Pinshane Y. Huang[3], Steven M. George[2], and J. Scott Bunch[1*]

[1] Department of Mechanical Engineering, University of Colorado, Boulder, CO 80309 USA

[2] Department of Chemistry and Biochemistry, University of Colorado, Boulder, CO 80309 USA

[3] School of Applied and Engineering Physics, Cornell University, Ithaca, NY 14853 USA

*email: jbunch@colorado.edu

**Abstract**

In this paper, a method is presented to create and characterize mechanically robust, free standing, ultrathin, oxide films with controlled, nanometer-scale thickness using Atomic Layer Deposition (ALD) on graphene. Aluminum oxide films were deposited onto suspended graphene membranes using ALD. Subsequent etching of the graphene left pure aluminum oxide films only a few atoms in thickness. A pressurized blister test was used to determine that these ultrathin films have a Young's modulus of 154 ± 13 GPa. This Young's modulus is comparable to much thicker alumina ALD films. This behavior indicates that these ultrathin two-dimensional films have excellent mechanical integrity. The films are also impermeable to standard gases suggesting they are pinhole-free. These continuous **ultrathin films are expected to enable new applications in fields such as thin film coatings, membranes and flexible electronics.**

Two-dimensional (2D) materials are promising nanomechanical structures [1,2]. Graphene, the best known and studied of this class of materials, boasts a high Young's modulus, intrinsic strength, gas impermeability, and excellent thermal and electrical conductivity [3–8]. There are numerous applications where flexible ultrathin insulating or oxide films are needed with comparable mechanical properties. The integration of graphene with other two dimensional (2D) or quasi-2D materials may also lead to new functional properties for the composite materials [9–13]. Currently, the range of ultra-thin materials is severely limited by the materials and length-scales that are accessible through thin film fabrication.

Mechanical and chemical exfoliation, as well as growth techniques such as chemical vapor deposition, can produce just a handful of ultra-thin layered materials [1,14–17]. As traditional materials approach ~ 1 nm film thicknesses, fabrication of freely suspended films is difficult due to stresses or significant voids in the films that destroy the mechanical integrity of the film. To overcome these problems, we use suspended graphene membranes as sacrificial supports to grow high quality ALD films and then remove the graphene to leave the ALD thin film. These experiments demonstrate that ALD on graphene offers a route to create free-standing, ultrathin, quasi-2D structures with atomically controlled thickness and mechanical properties comparable to their bulk counterparts [18–21].

Atomic layer deposition films are fabricated using a combination of deposition and etching using a suspended graphene support. The graphene provides an atomically-smooth growth surface that can easily be etched away. Graphene is mechanically exfoliated over predefined wells as illustrated in Fig. 1a. The graphene is then exposed to a trimethylaluminum (TMA) and $NO_2$ treatment that forms an adhesion layer for ALD nucleation [22–24]. Aluminum oxide ALD is subsequently grown using TMA/$H_2O$ doses [25,26] (see supporting online text). An example of such a graphene/ALD composite film after 7 cycles of alumina ALD is shown in Fig. 1b. High resolution cross-sectional transmission electron microscopy on such a graphene sample with a TMA/$NO_2$ adhesion layer followed by 5 cycles of TMA/$H_2O$ shows the ALD film to be amorphous and 2.8 ± 0.3 nm thick (see supporting online text). We then use oxidative etching of the underlying graphene support to leave only the thin alumina ALD film suspended over the predefined well as displayed in Fig. 1c. Oxidative etching is carried out in a 1 inch diameter tube furnace at 600° C with an $O_2$ gas flow of ~20-40 ccm for ~ 10 hours. This is sufficient to completely etch away the graphene. After the graphene is etched away, the film is no longer visible in the optical microscope, and Raman spectroscopy which shows no signs of a substantial D, G, and 2D peak in the etched samples is used to confirm the absence of graphene (see supporting online text).

A pressure difference is applied to the film using a previously reported method where slow diffusion through the $SiO_2$ substrate over pressurizes the film sealed microchamber [5,27]. An atomic force microscope (AFM) image of such an over pressurized suspended film in Fig. 1c is shown in Fig. 1d. The ALD film is bulged upward with a maximum deflection through the center of the film, $\delta = 261$ nm, and a radius, $a = 2.76$ μm, consistent with the radius of the predefined well. At increasing $\Delta p$, the film stretches further as $\delta$ increases as characterized in Fig. 1e. During AFM imaging, the bulge is stable suggesting a constant pressure difference and no significant leak rate of gas out of the microchamber, similar to previous results on graphene membranes [5]. This behavior implies that the aluminum oxide films are pinhole-free and impermeable to the nitrogen gas used for pressurization.

The deformation of the film follows [27,28]:

$$\Delta p = K(\upsilon) \frac{\delta^3}{a^4} Et \qquad (1)$$

where $E$ is Young's modulus, $t$ is the thickness of the film, and $K(\upsilon)$ is a constant that depends on the Poisson's ratio. For the case of aluminum oxide, $K(\upsilon = 0.24) = 3.35$. Figure 2a shows $K(\upsilon)\frac{\delta^3}{a^4}$ vs. $\Delta p$ for 18 pure alumina ALD films (graphene etched away) fabricated on an exfoliated graphene flake using 7 cycles of alumina ALD. The behavior of each film follows a line as expected from eq. (1). The average and standard deviation of the slope of these lines gives $Et = 250 \pm 12$ GPa-nm.

A similar measurement was performed for a number of different films formed using 4 – 15 cycles. The plot of $Et$ vs. number of ALD cycles is shown in Fig. 2b. A best fit line of the data gives a slope of $Et_{cycle} = 16.9 \pm 1.4$ GPa – nm with an intercept of $E_0 t_0 = 127.1 \pm 13.1$ GPa – nm. This non-zero intercept likely arises from the $Et$ value of the functionalization layer. This slope corresponds to $E_{ALD\ Al2O3} = 154 \pm 13$ GPa assuming an ALD growth rate of 0.11 nm/cycle [25,29]. This Young's modulus is comparable to previous measurements on much thicker (tens to hundreds of nm) alumina ALD films that have Young's moduli of 168 – 220 GPa [30–32]. Because the films are freely suspended, a mechanical support does not influence the mechanical properties of the ALD thin films. The high Young's modulus is remarkable considering our samples are 2 – 3 orders of magnitude thinner than previously measured ALD films.

The pressure induced-strain in the film can be used to tune the mechanical resonance frequency of the suspended films. Figure 3a demonstrates this behavior for a graphene/ALD composite film fabricated using 5 cycles of alumina ALD. The mechanical resonance is actuated and detected optically as previously reported [3,5]. We were unable to measure a resonance frequency for the pure alumina ALD films presumably due to the lack of optical reflectivity from these samples. The frequency first decreases and then increases as the film transitions from a bulged upward to a bulged downward state.

At sufficiently large pressures far from the minimum frequency, the frequency scales as $f^3 \alpha \Delta p$. The slope shows a dramatic decrease in frequency with the addition of alumina ALD cycles as shown in Fig. 3b. This behavior can be explained by the pressure-induced changes in the tension in a stretched circular film according to:

$$f^3 = 7 \times 10^{-3} \sqrt{\frac{K(\upsilon)Et}{a^4 \rho_A^3}} \Delta p \qquad (2)$$

where $\rho_A$ is the mass per unit area. From the slope of the lines in Fig. 3b and using $K(\upsilon)Et$ determined by a pressurized blister test on the composite ALD/graphene film (see supporting online text), we can determine $\rho_A$ of each suspended film before and after each ALD process. All samples showed an increase in $\rho_A$ with number of alumina ALD cycles as displayed in Fig. 3c. The first three cycles showed a larger increase in $\rho_A$ that may be related to the initial nucleation of alumina ALD. The finite $\rho_A$ before any ALD cycles is attributed to the additional mass from the adhesion layer.

Using the measured $\rho_A$, we can estimate the volume density of the ALD films, $\rho_V$, and the areal mass density of the adhesion layer, $\rho_{A\_ad}$. Because all the samples have an adhesion layer with an unknown $\rho_{A\_ad}$, we first determine $\rho_V$ from the slope of the lines in Fig. 3c for coatings after the nucleation treatment. This determination yields $\rho_V = 2.3 \pm 0.4$ g/cm$^3$ assuming an ALD growth rate of 0.11 nm/cycle [25]. (The anomalously large value at 12 cycles shown in blue was not used in calculating this average and standard deviation.) We then deduce $\rho_{A\_ad}$ from the measured $\rho_A$ using $\rho_{A\_ad} = \rho_A - \rho_V * N$, where $N$ is the number of alumina ALD cycles. This derivation yields an average value and standard deviation of $\rho_{A\_ad} = 1.4 \pm 0.3 * 10^{-7}$ g/cm$^2$. We can then determine $\rho_V$ for every ALD film in Fig. 3c. This procedure yields $\rho_V = 2.4 \pm 0.7$ g/cm$^3$ as shown in Fig. 3d. This density is comparable within experimental error to previous densities measured on thicker alumina ALD thin films of 3.0 g/cm$^3$ [33].

None of the 178 samples fabricated with less than 4 alumina ALD cycles were impermeable to N$_2$ gas after removal of the graphene as shown in Fig. 4a. However, the yield of impermeable films increased with number of ALD cycles and reached 85% for 15 ALD cycles as displayed in Fig. 4b. This behavior indicates that increasing the number of ALD cycles reduces pinholes or gas diffusion through the film.

For freely suspended films formed using only 5 cycles of the TMA/NO$_2$ nucleation treatment, AFM images of the films do not show voids (see supporting online text). This corroborates our measurements of a contribution from the adhesion layer to $E$ and $\rho_{A\_funct}$ in Fig. 2b and Fig. 3c. Future work will examine the dependence of the adhesion layer and its role in nucleating continuous pinhole-free ALD alumina thin film growth on graphene.

In conclusion, a new class of ultrathin films has been created based on aluminum oxide ALD on graphene. These films are mechanically robust, pinhole-free, and have ~nm thicknesses while still maintaining a Young's modulus comparable to their much thicker counterparts. The manufacturability, thickness control, and versatility of the ALD process means that materials and processing can be tailored to suit many applications where traditional silicon or graphene-based thin film mechanical devices fail to offer the needed functionality [34,35]. Furthermore, these films can be integrated with graphene or other nanomechanical structures to create multifunctional quasi-2D electromechanical structures.

ACKNOWLEDGEMENT: We thank Darren McSweeney, and Michael Tanksalvala for help with the resonance measurements, Rishi Raj for use of the Raman microscope, and Narasimha Boddetti, Jianliang Xiao, Martin L. Dunn, Victor Bright, Todd Murray, David Muller, Paul McEuen, and Yifu Ding for useful discussions. This work was supported by NSF Grants #0900832(CMMI: Graphene Nanomechanics: The Role of van der Waals Forces), #1054406(CMMI: CAREER: Atomic Scale Defect Engineering in Graphene Membranes), the DARPA Center on Nanoscale Science and Technology for Integrated Micro/Nano-Electromechanical Transducers (iMINT), the National Science Foundation (NSF) Industry/University Cooperative Research Center for Membrane Science, Engineering and Technology (MAST), and in part by the NNIN and the National Science

Foundation under Grant No. ECS-0335765. Electron Microscopy facilities were provided by the NSF through the Cornell Center for Materials Research (NSF DMR-1120296).Supporting Information Available: Experimental Methods, TEM Imaging of a Graphene/ALD Composite, Raman Spectrum Verifying the Etching of Graphene, Elastic Constants of Pure ALD Films, Initial Tension in Graphene and Graphene/ALD Composite Films, and Pure ALD Films from the Nucleation Treatment. This material is available free of charge via the internet at http://pubs.acs.org.

**References**

(1) Novoselov, K. S.; Jiang, D.; Schedin, F.; Booth, T. J.; Khotkevich, V. V.; Morozov, S. V.; Geim, A. K. *Proceedings of the National Academy of Sciences of the United States of America* **2005**, *102*, 10451-10453.

(2) Geim, A. K. *Science* **2009**, *324*, 1530-1534.

(3) Bunch, J. S.; van der Zande, A. M.; Verbridge, S. S.; Frank, I. W.; Tanenbaum, D. M.; Parpia, J. M.; Craighead, H. G.; McEuen, P. L. *Science* **2007**, *315*, 490-493.

(4) Lee, C.; Wei, X.; Kysar, J. W.; Hone, J. *Science* **2008**, *321*, 385-388.

(5) Bunch, J. S.; Verbridge, S. S.; Alden, J. S.; Zande, A. M. V. D.; Parpia, J. M.; Craighead, H. G.; McEuen, P. L. *Nano Letters* **2008**, *8*, 2458-2462.

(6) Novoselov, K. S.; Geim, A. K.; Morozov, S. V.; Jiang, D.; Katsnelson, M. I.; Grigorieva, I. V.; Dubonos, S. V.; Firsov, A. A. *Nature* **2005**, *438*, 197-200.

(7) Zhang, Y. B.; Tan, Y. W.; Stormer, H. L.; Kim, P. *Nature* **2005**, *438*, 201-204.

(8) Balandin, A. A.; Ghosh, S.; Bao, W.; Calizo, I.; Teweldebrhan, D.; Miao, F.; Lau, C. N. *Nano Letters* **2008**, *8*, 902-907.

(9) R., D.; F., Y.; MericI.; LeeC.; WangL.; SorgenfreiS.; WatanabeK.; TaniguchiT.; KimP.; L., S.; HoneJ. *Nat Nano* **2010**, *5*, 722-726.

(10) Williams, J. R.; DiCarlo, L.; Marcus, C. M. *Science* **2007**, *317*, 638-641.
8

(11)  Rogers, J. A.; Lagally, M. G.; Nuzzo, R. G. *Nature* **2011**, *477*, 45-53.

(12)  Jen, S.-H.; Bertrand, J. A.; George, S. M. *Journal of Applied Physics* **2011**, *109*, 84305-84311.

(13)  Huang, P. Y.; Kurasch, S.; Srivastava, A.; Skakalova, V.; Kotakoski, J.; Krasheninnikov, A. V.; Hovden, R.; Mao, Q.; Meyer, J. C.; Smet, J.; Muller, D. A.; Kaiser, U. *Nano Letters* **2012**, *12*, 1081-1086.

(14)  Coleman, J. N.; Lotya, M.; O'Neill, A.; Bergin, S. D.; King, P. J.; Khan, U.; Young, K.; Gaucher, A.; De, S.; Smith, R. J.; Shvets, I. V.; Arora, S. K.; Stanton, G.; Kim, H.-Y.; Lee, K.; Kim, G. T.; Duesberg, G. S.; Hallam, T.; Boland, J. J.; Wang, J. J.; Donegan, J. F.; Grunlan, J. C.; Moriarty, G.; Shmeliov, A.; Nicholls, R. J.; Perkins, J. M.; Grieveson, E. M.; Theuwissen, K.; McComb, D. W.; Nellist, P. D.; Nicolosi, V. *Science* **2011**, *331*, 568-571.

(15)  Li, X.; Cai, W.; An, J.; Kim, S.; Nah, J.; Yang, D.; Piner, R.; Velamakanni, A.; Jung, I.; Tutuc, E.; Banerjee, S. K.; Colombo, L.; Ruoff, R. S. *Science* **2009**, *324*, 1312-1314.

(16)  Zhan, Y.; Liu, Z.; Najmaei, S.; Ajayan, P. M.; Lou, J. *Small* **2012**, n/a--n/a.

(17)  Suk, J. W.; Murali, S.; An, J.; Ruoff, R. S. *Carbon* **2012**, *50*, 2220-2225.

(18)  Leskelä, M.; Ritala, M. *Thin Solid Films* **2002**, *409*, 138-146.

(19)  Leskelä, M.; Ritala, M. *Angewandte Chemie International Edition* **2003**, *42*, 5548-5554.

(20)  Elam, J. W.; Groner, M. D.; George, S. M. *Review of Scientific Instruments* **2002**, *73*, 2981-2987.

(21)  Ritala, M.; Leskelä, M.; Dekker, J.-P.; Mutsaers, C.; Soininen, P. J.; Skarp, J. *Chemical Vapor Deposition* **1999**, *5*, 7-9.

(22)  Cavanagh, A. S.; Wilson, C. a; Weimer, A. W.; George, S. M. *Nanotechnology* **2009**, *20*, 255602.

(23)  Farmer, D. B.; Gordon, R. G. *Nano Letters* **2006**, *6*, 699-703.

(24)  Wang, X.; Tabakman, S. M.; Dai, H. *Journal of the American Chemical Society* **2008**, *130*, 8152-8153.

(25)  Ott, A. W.; Klaus, J. W.; Johnson, J. M.; George, S. M. *Thin Solid Films* **1997**, *292*, 135-144.

(26) Dillon, A. C.; Ott, A. W.; Way, J. D.; George, S. M. *Surface Science* **1995**, *322*, 230-242.

(27) Koenig, S. P.; Boddeti, N. G.; Dunn, M. L.; Bunch, J. S. *Nat Nano* **2011**, *6*, 543-546.

(28) Hencky, H. *Zeitschrift fur Mathematik und Physik* **1915**, *63*, 311-317.

(29) Ott, A. W.; McCarley, K. C.; Klaus, J. W.; Way, J. D.; George, S. M. *Applied Surface Science* **1996**, *107*, 128-136.

(30) Tripp, M. K.; Stampfer, C.; Miller, D. C.; Helbling, T.; Herrmann, C. F.; Hierold, C.; Gall, K.; George, S. M.; Bright, V. M. *Sensors and Actuators A: Physical* **2006**, *130-131*, 419-429.

(31) Tapily, K.; Jakes, J. E.; Stone, D. S.; Shrestha, P.; Gu, D.; Baumgart, H.; Elmustafa, A. A. *Journal of The Electrochemical Society* **2008**, *155*, H545-H551.

(32) Miller, D. C.; Foster, R. R.; Jen, S.-H.; Bertrand, J. A.; Cunningham, S. J.; Morris, A. S.; Lee, Y.-C.; George, S. M.; Dunn, M. L. *Sensors and Actuators A: Physical* **2010**, *164*, 58-67.

(33) Groner, M. D.; Fabreguette, F. H.; Elam, J. W.; George, S. M. *Chemistry of Materials* **2004**, *16*, 639-645.

(34) Davidson, B. D.; Seghete, D.; George, S. M.; Bright, V. M. *Sensors and Actuators A: Physical* **2011**, *166*, 269-276.

(35) Yoneoka, S.; Lee, J.; Liger, M.; Yama, G.; Kodama, T.; Gunji, M.; Provine, J.; Howe, R. T.; Goodson, K. E.; Kenny, T. W. *Nano letters* **2012**, *12*, 683-6.

**Figure Captions**

1. **(a)** Schematic of a graphene membrane before atomic layer deposition (ALD). **(b)** (upper) Optical image of an exfoliated graphene flake with 7 cycles of alumina ALD. (lower) side view schematic of this graphene/ALD composite. **(c)** Optical image of a pure alumina film after graphene is etched away. (lower) side view schematic of this pure ALD film. **(d)** (upper) Atomic force microscope image of a pressurized 7 cycle pure alumina ALD film with $\Delta p$ = 278 kPa. This film corresponds to the film boxed in red in (b) and (c). **(e)** Deflection vs. position through the center of the film in (d) at different $\Delta p$.

2. **(a)** $K(v)z^3/a^4$ versus $\Delta p$ for 18 pure ALD films with 7 cycles of alumina ALD. Colored lines are best fits to each sample. The average and standard deviation of all the slopes corresponds to $Et = 250 \pm 12$ GPa-nm. **(b)** $Et$ vs. # of cycles for all the pure ALD films measured. The standard deviation is shown as error bars. The solid line is a best fit to the data and corresponds to $Et_{cycle} = 16.9 \pm 1.4$ GPa – nm with an intercept of $E_0 t_0 = 127.1 \pm 13.1$ GPa – nm. This corresponds to $E_{ALD\ Al2O3} = 154 \pm 13$ GPa assuming a thickness gain per cycle of $t_{cycle} = 0.11$ nm.

3. **(a)** Mechanical resonant frequency vs. $p_{ext}$ for a graphene /ALD composite film with 5 cycles of alumina ALD. (insets) Schematic of the film at different $\Delta p$. **(b)** Frequency$^3$ vs. $p_{ext}$ for a single graphene/ALD composite film with 0, 4, 9, 12 cycles of alumina ALD. **(c)** Areal mass density $\rho_A$ vs. number of cycles for all the graphene/ALD composites measured. **(d)** Histogram of volume mass density $\rho_V$ for the alumina ALD films. The black line is a Gaussian fit to the data.

4. **(a)** (black) Number of all pure ALD films fabricated in this study vs. number of ALD coating cycles. (red) Number of pure ALD films that hold $N_2$ gas from that sample batch **(b)** Percentage yield vs. # of cycles for all the pure ALD films fabricated.

**Figure**s

Fig.1

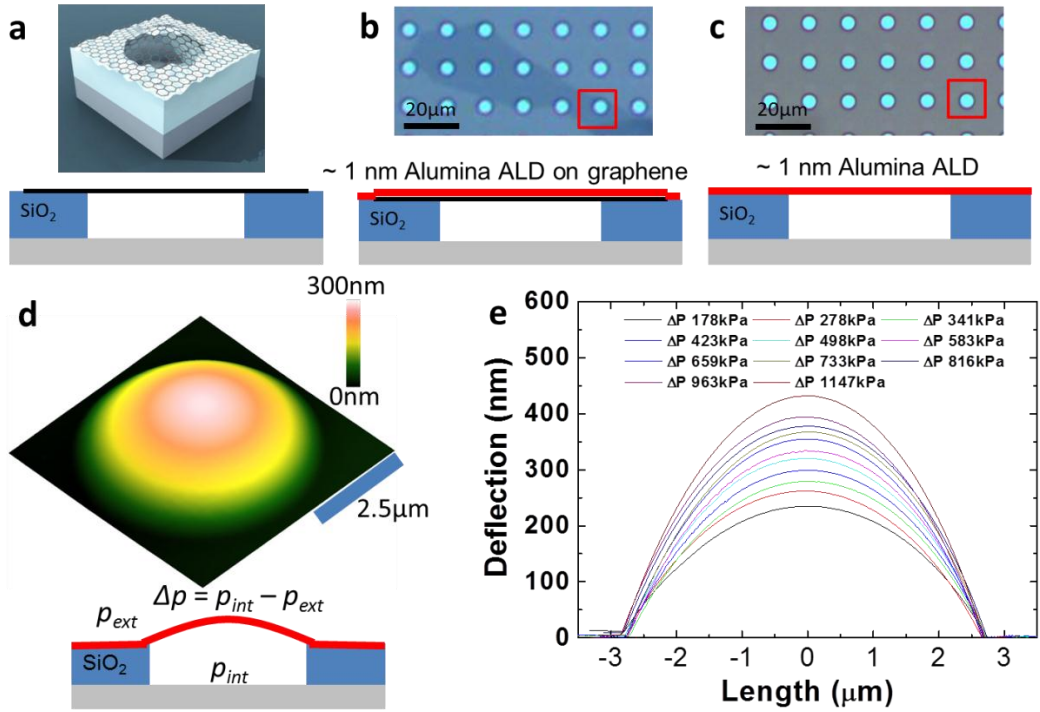

Fig.2

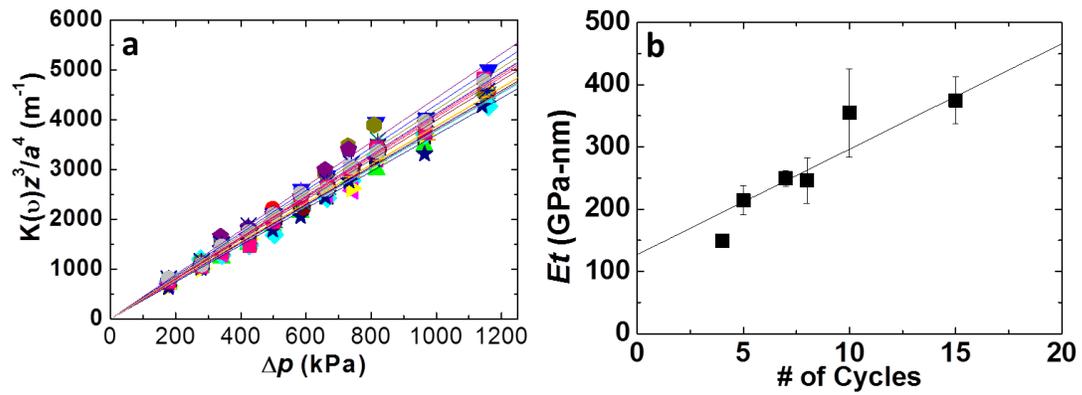

Fig.3

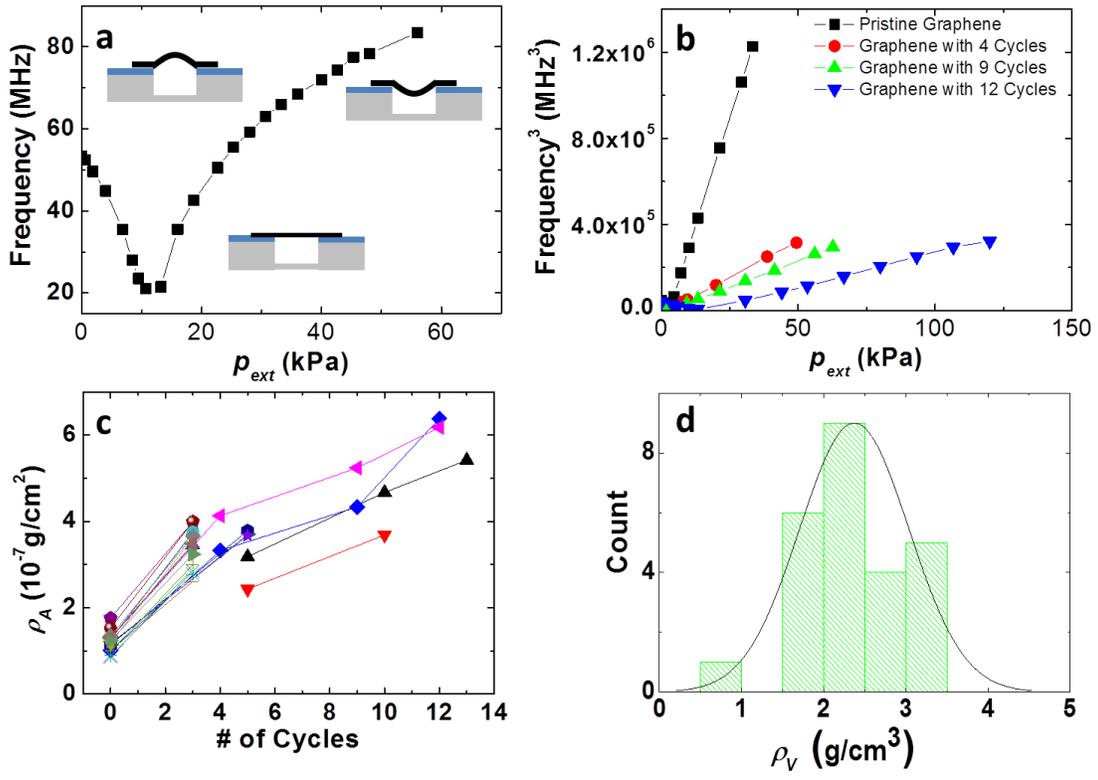

Fig.4

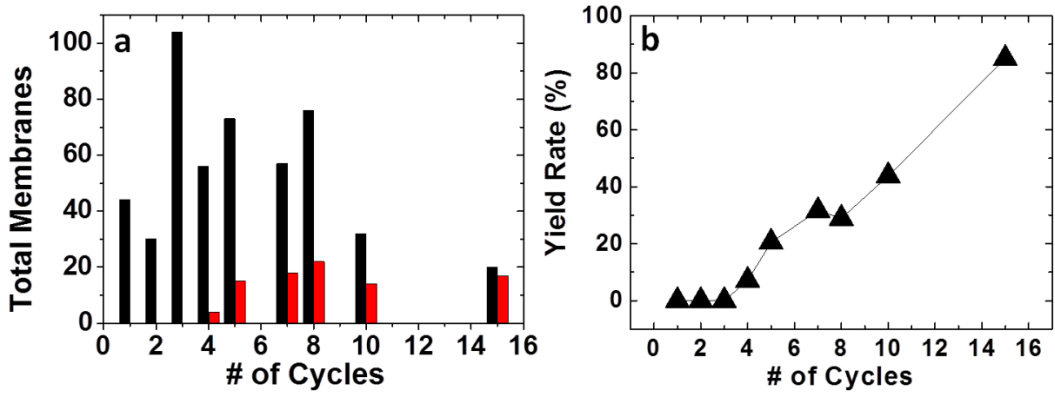

# Supplementary Information

# Ultrathin Oxide Films by Atomic Layer Deposition on Graphene

Luda Wang[1], Jonathan J. Travis[2], Andrew S. Cavanagh[2], Xinghui Liu[1], Steven P. Koenig[1], Pinshane Y. Huang[3], Steven M. George[2], and J. Scott Bunch[1*]

[1] *Department of Mechanical Engineering, University of Colorado, Boulder, CO 80309 USA*

[2] *Department of Chemistry and Biochemistry, University of Colorado, Boulder, CO 80309 USA*

[3] *School of Applied and Engineering Physics, Cornell University, Ithaca, NY 14853 USA*

*email: jbunch@colorado.edu

**Experimental Methods**

Graphene is deposited using mechanical exfoliation over pre-defined etched wells similar to previously described methods (see Fig. 1a) [1–3]. A series of wells with ~5-8 μm diameters are defined by photolithography on a silicon wafer with 90 nm of thermally grown silicon oxide. Dry plasma etching ($CF_4 + O_2$, followed by $SF_6$) is used to etch wells that are 500 nm – 3 μm and the "scotch tape" method is used to deposit graphene[1].

Atomic layer deposition on the graphene is performed in a homebuilt reactor following a recipe similar to one previously reported for ALD alumina growth on carbon nanotubes [4,5]. The samples are placed inside the ALD reactor, pumped down to ~30 mTorr, and held at 180°C for 30 min - 1 hour before beginning the reaction. The reactor is purged with 20 sequences of argon purging before performing the $NO_2$/TMA nucleation treatment. Each argon purge involved dosing argon to 1 Torr for 60 seconds and then pumping for 60 seconds.

The nucleation treatment involves a dose of $NO_2$ to 1 Torr for 60 sec followed by pumping for 60 seconds. Subsequently, a dose of TMA to 1 Torr for 60 sec is applied, followed by pumping for 60 seconds. This process is repeated 10 times. After forming this adhesion layer, ALD of alumina is performed by cycling $TMA/H_2O$ doses as follows: dose TMA to 1 Torr for 60 sec, pump for 60 sec, then dose Argon at 20 Torr for 60 sec and pump for 60 seconds 5 times, dose $H_2O$ to 1 Torr for 60 sec, pump for 60 sec, then dose Argon at 20 Torr for 60 sec and pump for 60 seconds 5 times. This represents 1 cycle of ALD. After the $TMA/H_2O$ cycles are complete, the reactor is purged again with argon and then the samples are removed. All reactions were performed at 180 ($\pm$ 0.5) °C.

### TEM Imaging of a Graphene/ALD Composite

Cross-sections of alumina/graphene devices on $SiO_2$ substrates were prepared using a focused ion beam lift-out. Before cross-sectioning, the samples were coated with ~20 nm of amorphous carbon followed by a thick platinum layer to protect the sample surfaces. The samples were imaged using a 200 kV electron beam in a FEI Technai-F-20 TEM/STEM. The composition of each layer was verified with electron energy-loss spectroscopy. TEM/STEM was used to image the alumina that was formed on graphene after 10 cycles of $TMA/NO_2$ nucleation treatment followed by 5 cycles of $TMA/H_2O$ ALD deposition. From these images, the observed alumina layer was amorphous and 2.8 $\pm$ 0.3 nm thick.

## Raman Spectrum Verifying the Etching of Graphene

Raman spectroscopy was used to confirm the removal of graphene after etching[1]. Figure S2a shows Raman spectroscopy of a graphene/ALD composite film before etching of graphene. The Raman spectrum shows the G and 2D peaks that are characteristic of a bilayer graphene membrane. After etching, the G, D, and 2D peaks are not detectable in the Raman spectrum confirming that all the graphene was etched away leaving only the suspended pure alumina ALD film (Fig. S2b). Note that the vertical scales in Fig. S2a and S2b are different. Figure S3 shows a comparison of the Raman spectrum on the same vertical scale before and after etching. This wave number range includes the peaks arising from the oxidized silicon substrate and a comparison of these peak sizes serves as a calibration of laser intensity between the 2 measurements. The lack of a visible D, G, and 2D peak in Fig. S3b confirms the successful oxidative etching and removal of graphene.

## Elastic Constants of Pure ALD Films

A method identical to one previously used to determine the elastic constants of pressurized graphene membranes was used to determine the elastic constants of the pure ALD films[2]. Figure S4 shows additional blister test data for the other pure ALD films fabricated and tested in this study. All of the films shown in one plot are from the same ALD coating on multiple graphene membrane supports. The *Et* values extracted from the slopes were plotted in Fig. 1b of the main text.

## Initial Tension in Graphene and Graphene/ALD Composite Films

Even with no applied pressure difference across the films, the frequency of these nanomechanical resonators still behave as stretched membranes[3]. This is illustrated by the high resonant frequencies exhibited by the membranes even when no pressure difference exists across the membranes or $p_{int} = p_{ext}$. Neglecting the bending rigidity, the fundamental frequency of a clamped circular membrane under uniform tension, $S_0$, is given by:

$$f_0 = \frac{2.404}{2\text{Å}} \cdot \sqrt{\frac{S_0}{ma^2}} \tag{S1}$$

The initial surface tension in the membranes can be deduced by measuring the resonant frequency of the membranes when no pressure difference exists across the membrane. For the graphene membrane resonator before ALD deposition, this corresponds to a uniform tension of $S_0 = 0.073 \pm 0.041$ N/m (Fig. S5a). After ALD film deposition, the uniform tension is $S_0 = 0.21 \pm 0.13$ N/m (Fig. S5b). This increase in uniform tension indicates that there is a significant increase in the intrinsic stress in the membranes. Future work will seek to understand the origin of this increased tension in the composite membranes to determine whether this intrinsic stress is a result of stress in the pure ALD films or arises from the composite nature of the films.

## Pure ALD Films from the Nucleation Treatment

The thinnest suspended pure alumina ALD film fabricated is shown in Figure S6. This film had only 4 cycles of the $NO_2$/TMA nucleation treatment applied to the graphene. Subsequent etching of the graphene support leaves a continuous and smooth film. The film has a few small voids visible by AFM. Figure S7 shows optical and AFM images for a film made from only 5 cycles of the $NO_2$/TMA nucleation treatment. This film is also continuous and smooth and there are no pinholes or small voids visible by AFM.

**Supplementary References**

(1) Novoselov, K. S.; Jiang, D.; Schedin, F.; Booth, T. J.; Khotkevich, V. V.; Morozov, S. V.; Geim, A. K. *Proceedings of the National Academy of Sciences of the United States of America* **2005**, *102*, 10451-10453.

(2) Bunch, J. S.; Verbridge, S. S.; Alden, J. S.; Zande, A. M. V. D.; Parpia, J. M.; Craighead, H. G.; McEuen, P. L. *Nano Letters* **2008**, *8*, 2458-2462.

(3) Koenig, S. P.; Boddeti, N. G.; Dunn, M. L.; Bunch, J. S. *Nat Nano* **2011**, *6*, 543-546.

(4) Cavanagh, A. S.; Wilson, C. a; Weimer, A. W.; George, S. M. *Nanotechnology* **2009**, *20*, 255602.

(5) Farmer, D. B.; Gordon, R. G. *Nano Letters* **2006**, *6*, 699-703.

## Supplementary Figure Captions

**Figure S1**

Bright-field TEM image of a cross-section of supported alumina ALD film on 5-layer graphene supported on silicon oxide. The amorphous alumina layer is 2.8 ± 0.3 nm thick.

**Figure S2**

**(a)** Raman spectrum for one of the graphene/ALD composite films in Fig. 1b. **(b)** A representative Raman spectrum on one of the pure alumina ALD films in Fig. 1C. Note that the vertical scales are different.

**Figure S3**

**(a)** Raman spectrum showing the full spectrum of the data in Fig, S2a **(b)** Raman spectrum showing the full spectrum of the data in Fig, S2b.

**Figure S4**

K($1/2$ $^{-3}$/$a^4$ versus •p for (a) 5 pure ALD films with 8 cycles of alumina ALD. The average and standard deviation of all the slopes corresponds to $Et = 213 \pm 12$ GPa-nm. (b) 1 pure ALD films with 5 cycles of alumina ALD. The slope is a best fit line and corresponds to $Et = 179 \pm 6$ GPa-nm. (c) 8 pure ALD films with 5 cycles of alumina ALD. The average and standard deviation of all the slopes corresponds to $Et = 219 \pm 21$ GPa-nm. (d) 5 pure ALD films with 8 cycles of alumina ALD. The average and standard deviation of all the slopes corresponds to $Et = 280 \pm 12$ GPa-nm. (e) 3 pure ALD films with 10 cycles of alumina ALD. The average and standard deviation of all the slopes corresponds to $Et = 355 \pm 71$ GPa-nm. (f) 9 pure ALD films with 15 cycles of alumina ALD. The average and standard deviation of all the slopes corresponds to $Et = 375 \pm 38$ GPa-nm.

**Figure S5**

Histogram of initial tension in **(a)** pristine graphene membranes and **(b)** graphene/ALD composite membranes

**Figure S6**

**(a)** Optical image of a graphene flake with 4 cycles of $NO_2$/TMA. **(b)** Optical image after etching away the graphene **(c)** Atomic force microscope image corresponding to the red box in **(b)** of the pure alumina ALD film (scale bar = 2.5 µm)

**Figure S7**

(a) Optical image of a graphene flake with 5 cycles of NO$_2$/TMA. (b) Optical image after etching away the graphene (c) Atomic force microscope image corresponding to the red box in (b) of the pure alumina ALD film (scale bar = 2.5 µm)

# Figures

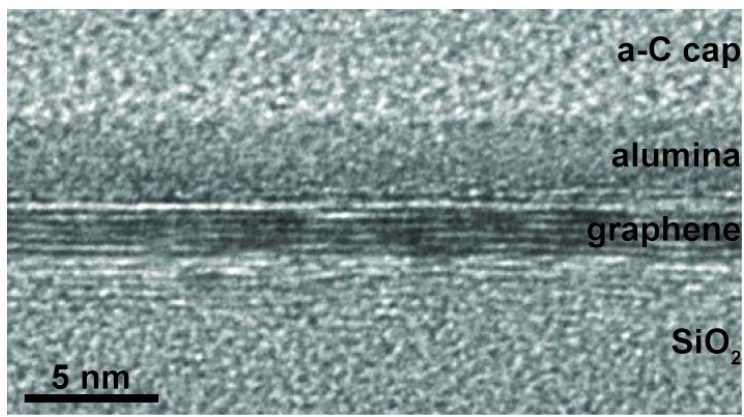

Figure S1

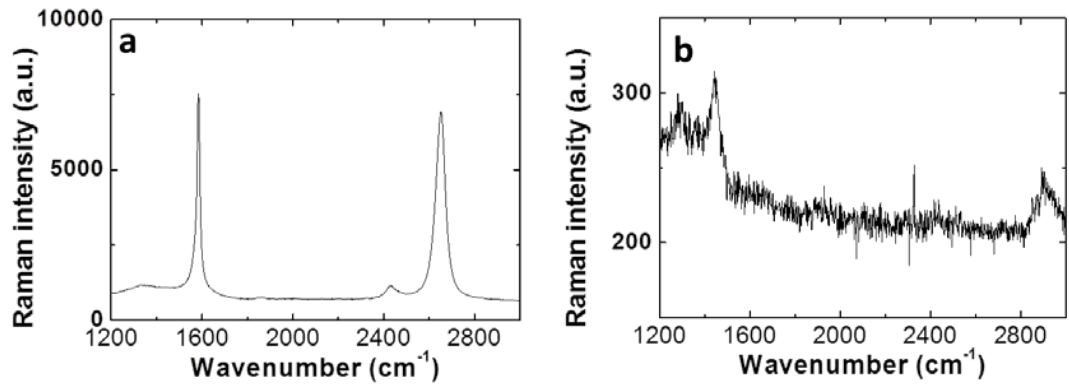

Figure S2

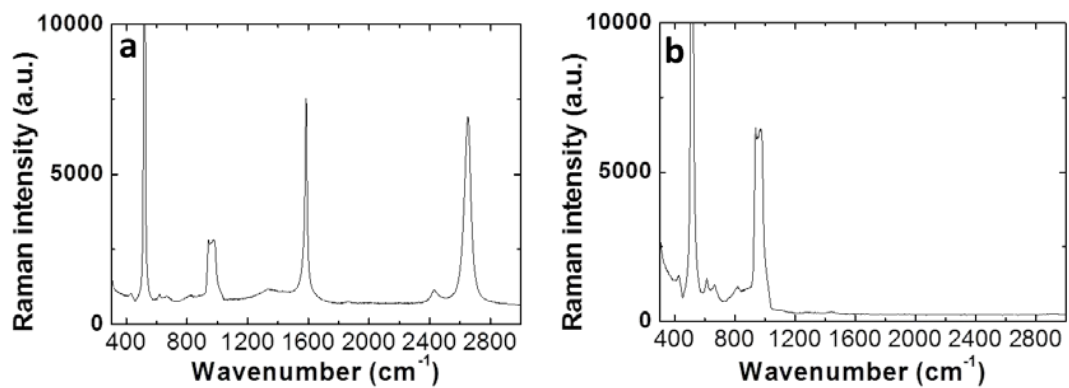

Figure S3

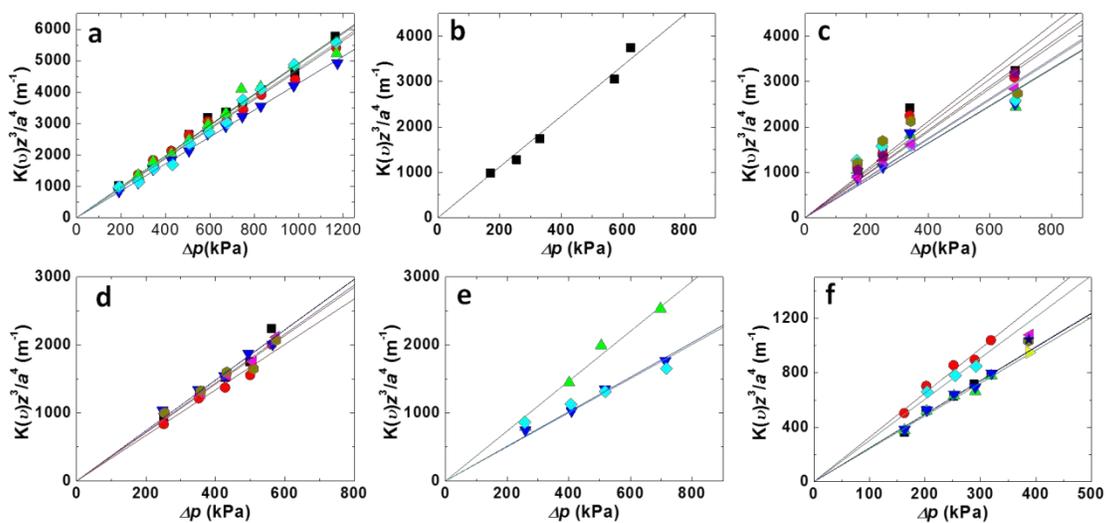

Figure S4

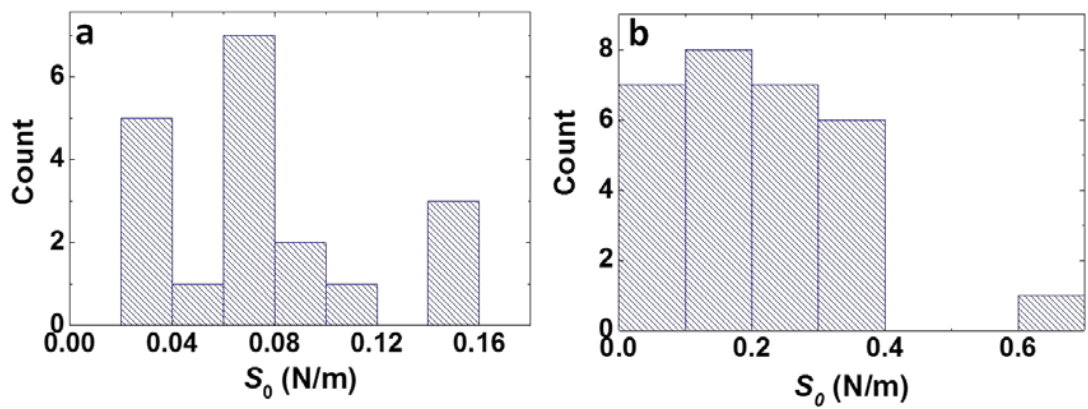

Figure S5

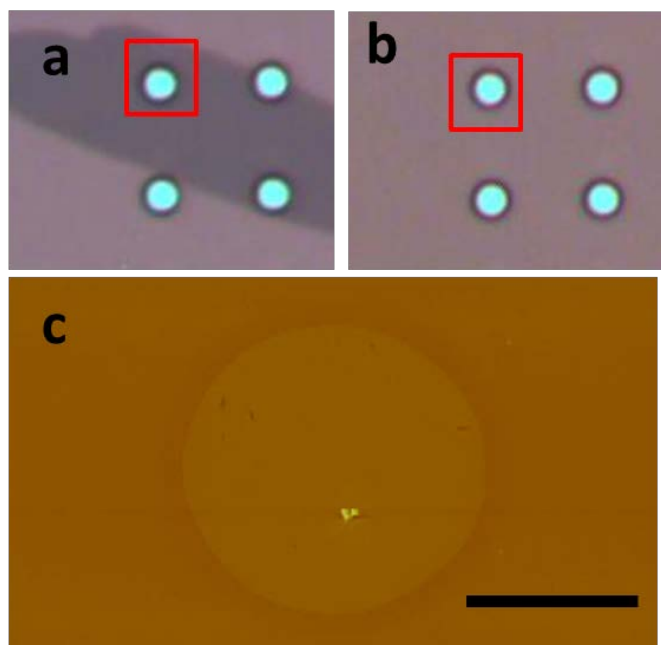

Figure S6

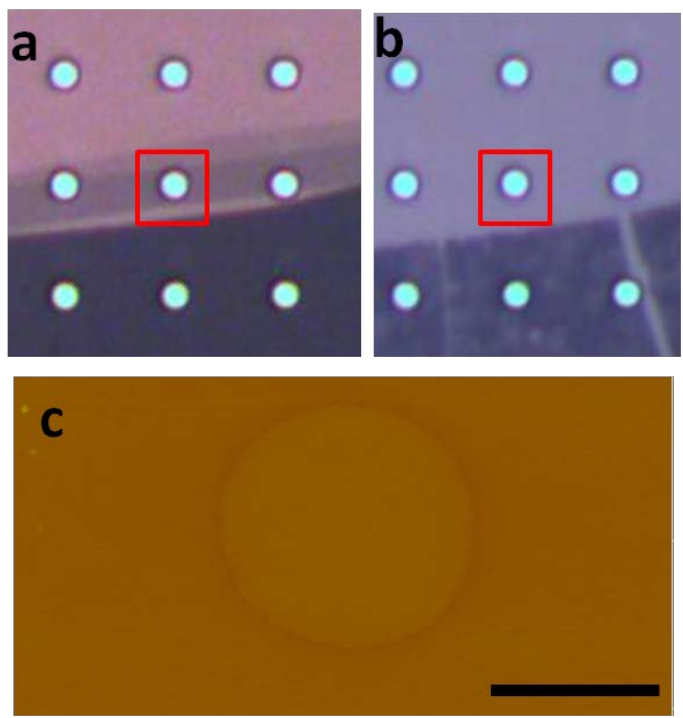

Figure S7